\documentclass[runningheads]{llncs}

\usepackage[T1]{fontenc}
\usepackage{graphicx}
\usepackage{amsmath,amssymb,amsfonts} 
\usepackage{booktabs}        
\usepackage{siunitx}         
\usepackage{array}
\usepackage{makecell}
\usepackage{multirow}
\usepackage{cite}
\usepackage{xcolor}
\usepackage{colortbl}        
\usepackage{hyperref}
\usepackage[ruled,vlined]{algorithm2e}

\usepackage{tikz}

\begin{document}

\title{Topology-Guided Biomechanical Profiling: A White-Box Framework for Opportunistic Screening of Spinal Instability on Routine CT}

\titlerunning{TGBP}

\author{
Zanting Ye\inst{1,7} \and
Xuanbin Wu\inst{1} \and
Guoqing Zhong\inst{2} \and
Shengyuan Liu\inst{3} \and
Jiashuai Liu\inst{4} \and
Ge Song\inst{1} \and
Zhisong Wang\inst{5} \and
Jing Hao\inst{6} \and
Xiaolong Niu\inst{1} \and
Yefeng Zheng\inst{7} \and
Yu Zhang\inst{8}\thanks{Corresponding author. Email: luck\_2001@126.com} \and
Lijun Lu\inst{1}\thanks{Corresponding author. Email: ljlubme@gmail.com}
}

\authorrunning{Ye et al.}

\institute{
Southern Medical University, Guangzhou, China \and
The Affiliated Cancer Hospital of Zhengzhou University, Zhengzhou, China \and
The Chinese University of Hong Kong, Hong Kong, China \and
Xi'an Jiaotong University, Xi'an, China \and
Northwestern Polytechnical University, Xi'an, China \and
The University of Hong Kong, Hong Kong, China \and
Westlake University, Hangzhou, China \and
Guangdong Provincial People's Hospital, Guangzhou, China
}
\maketitle              



\begin{abstract}
Routine oncologic computed tomography (CT) presents an ideal opportunity for screening spinal instability, yet prophylactic stabilization windows are frequently missed due to the complex geometric reasoning required by the Spinal Instability Neoplastic Score (SINS). Automating SINS is fundamentally hindered by metastatic osteolysis, which induces topological ambiguity that confounds standard segmentation and black-box AI. We propose Topology-Guided Biomechanical Profiling (TGBP), an auditable white-box framework decoupling anatomical perception from structural reasoning. TGBP anchors SINS assessment on two deterministic geometric innovations: (i) canal-referenced partitioning to resolve posterolateral boundary ambiguity, and (ii) context-aware morphometric normalization via covariance-based oriented bounding boxes (OBB) to quantify vertebral collapse. Integrated with auxiliary radiomic and large language model (LLM) modules, TGBP provides an end-to-end, interpretable SINS evaluation. Validated on a multi-center, multi-cancer cohort ($N=482$), TGBP achieved 90.2\% accuracy in 3-tier stability triage. In a blinded reader study ($N=30$), TGBP significantly outperformed medical oncologists on complex structural features ($\kappa=0.857$ vs.\ $0.570$) and prevented compounding errors in Total Score estimation ($\kappa=0.625$ vs.\ $0.207$), democratizing expert-level opportunistic screening.

\end{abstract}

\begin{figure}[t]
    \centering
    \includegraphics[width=0.9\textwidth]{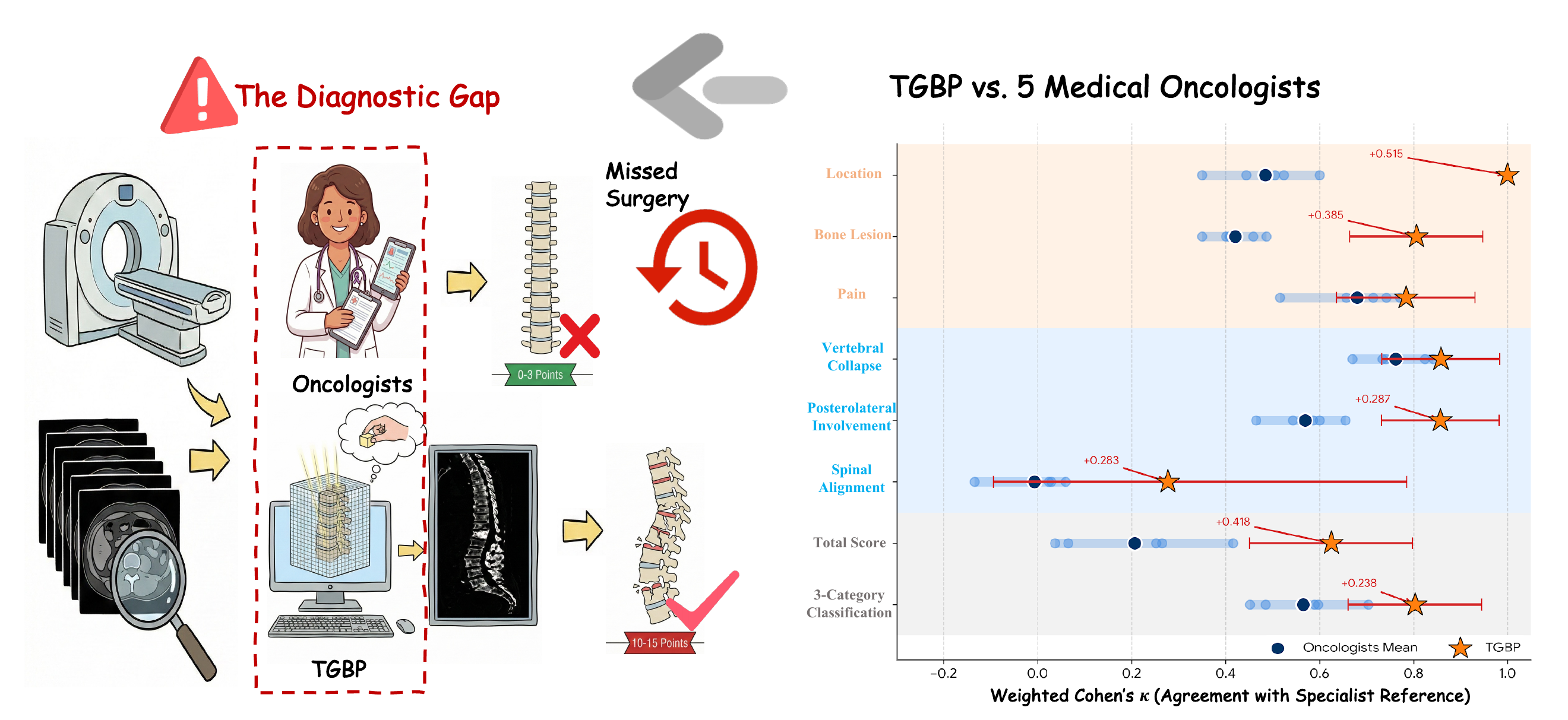} 
    \caption{Motivation of TGBP. A reader study ($N=30$) exposes a diagnostic gap where medical oncologists struggle with complex SINS criteria, leading to missed surgical windows.}
    \label{fig:motivation}
\end{figure}

\section{Introduction}

Spinal metastases affect up to 40\% of advanced cancer patients, necessitating vigilant surveillance \cite{chow2007palliative,bartels2008spinal}. Routine thoracoabdominal computed tomography (CT) frequently captures these lesions, creating an opportunity for \emph{opportunistic screening} \cite{costelloe2009imaging,pickhardt2020automated}. The Spinal Instability Neoplastic Score (SINS) \cite{fisher2010novel,fourney2011spinal} is the standard clinical triage framework. However, its complex structural criteria—such as morphometric collapse and posterolateral involvement—demand orthopedic spatial reasoning. Non-specialists often struggle with these assessments, creating a severe diagnostic gap (Fig.~\ref{fig:motivation}). Our reader study ($N=30$) reveals that individual variance among medical oncologists cascades into a catastrophic compounding error in Total SINS estimation (mean $\kappa=0.207$), leading to delayed prophylactic surgeries \cite{cole2008metastatic}.

\textbf{Computational Challenges and Prior Work.} Automating SINS is hindered by metastatic osteolysis. Conventional deep learning architectures \cite{wasserthal2023totalsegmentator, ma2023medsam} and computer-aided diagnosis systems for osteoporotic fractures \cite{peng2024study} implicitly assume intact cortical boundaries. Osteolytic destruction violates this geometric prior, rendering structural partitioning ill-posed \cite{hu2019topology,wen2025topology}. Furthermore, end-to-end black-box models map image patches directly to risk scores \cite{liu2024learning,guo2024radiographic}. By bypassing explicit boundaries, they fail to expose the visualizable geometric evidence required to build clinical trust. Recent LLM and VLM based approaches \cite{chan2025llm,ye2026unveiling,jiang2025hulu,xu2025lingshu,sellergren2025medgemma} process radiology images and reports but cannot extract primary spatial evidence. No prior work has delivered an end-to-end system capable of holistic SINS profiling with auditable geometric reasoning.

\textbf{Proposed Framework.} We propose Topology-Guided Biomechanical Profiling (TGBP), an auditable white-box framework acting as an expert concurrent reader. By confining deep learning strictly to upstream anatomical parsing, TGBP ensures that downstream risk stratification is driven by transparent geometric invariants. Crucially, while all reported metrics evaluate the system in a strictly automated mode to demonstrate algorithmic robustness, TGBP operationalizes clinical accountability via an immersive Virtual Reality (VR) supervisory port, facilitating the expert intervention intrinsically essential for trustworthy AI. Our main contributions are:
\begin{enumerate}
    \item We resolve ill-posed boundary problems under metastatic osteolysis via a canal-referenced topological partitioning.
    \item We introduce context-aware morphometric normalization using covariance-based bounding boxes (OBBs) to extract auditable collapse evidence robust to scoliotic deformity.
    \item Validated on a multi-center, multi-cancer cohort ($N=482$), TGBP accurately predicts clinical stability (up to 90.2\% accuracy) and significantly outperforms medical oncologists, successfully bridging the diagnostic gap.
\end{enumerate}

\section{Methodology}
\begin{figure}[t]
    \centering
    \includegraphics[width=\textwidth]{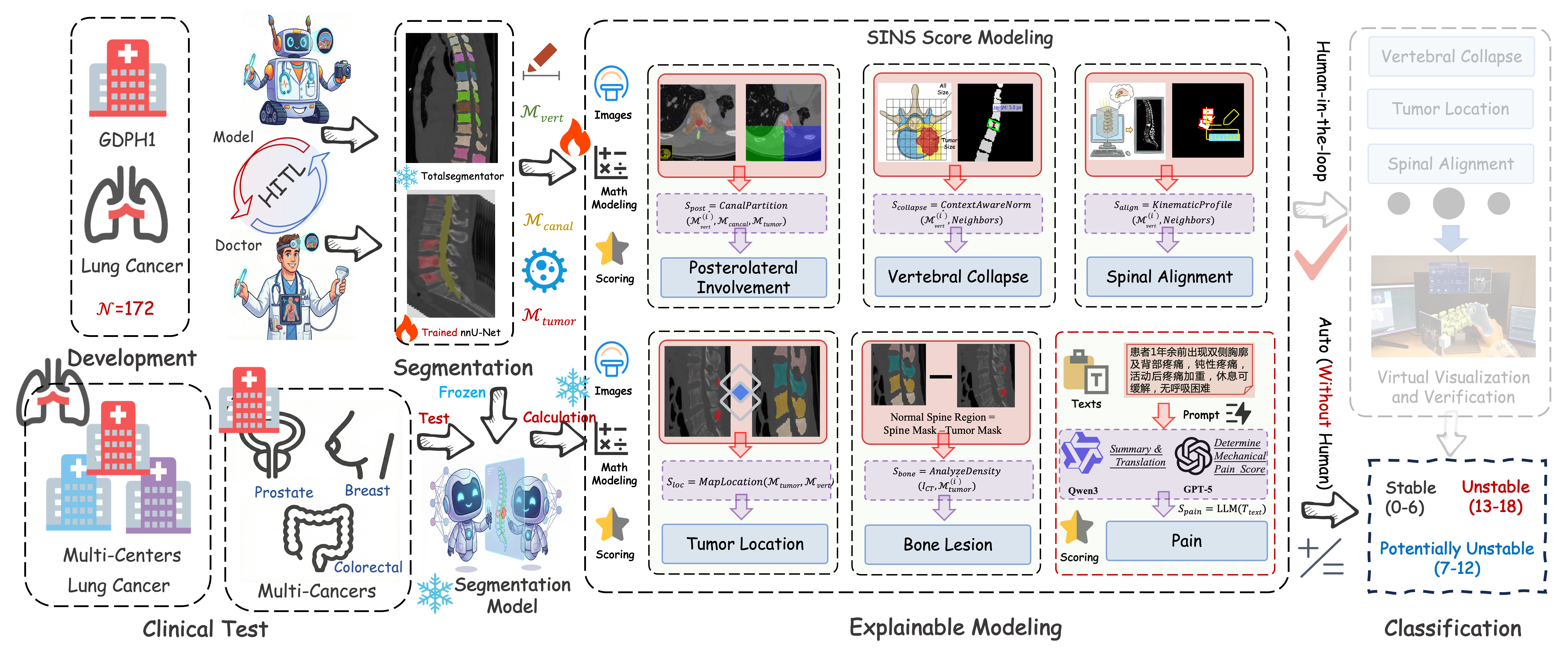} 
    \caption{Overview of the TGBP framework. \textbf{Left (Segmentation):} A segmentation backbone trained via HITL strategy is frozen for downstream deployment. \textbf{Middle (Explainable Modeling):} The white-box engine translates anatomical masks and clinical text into explicit SINS criteria via transparent geometry and LLMs. \textbf{Right (Triage \& VR Integration):} Profiles are aggregated for 3-tier stability classification. Notably, TGBP supports both fully automated screening and a semi-automated VR mode for immersive expert verification and correction.}
    \label{fig:workflow}
\end{figure}

TGBP operationalizes the multifaceted SINS criteria (Fig. \ref{fig:workflow}, Algorithm \ref{alg:tgbp}), achieving strong interpretability by decoupling perception from reasoning. Specifically, an anatomical parsing module trained via a rigorous Human-in-the-Loop (HITL) strategy is frozen for automated downstream deployment. The subsequent risk stratification is executed exclusively through an explainable geometric engine. By bypassing black-box heuristics in favor of transparent mathematics, TGBP directly mirrors orthopedic cognitive processes to output a fully auditable 3-tier stability classification. Beyond fully automated triage, TGBP features a semi-automated Virtual Reality (VR) mode for immersive expert correction.

\begin{algorithm}[t]
\caption{TGBP Pipeline}\label{alg:tgbp}
\SetAlgoLined
\KwIn{CT Volume $I_{CT}$, Clinical Notes $T_{text}$, Mode $\mathcal{V} \in \{\text{Auto}, \text{VR}\}$}
\KwOut{Total SINS Score $S_{total}$, 3-Tier Stability Class $C_{tier}$}

\tcp{Step 1: Anatomical Parsing \& Indexing}
$\mathcal{M}_{vert} \leftarrow \text{TotalSegmentator}(I_{CT})$\; 
$\mathcal{M}_{canal}, \mathcal{M}_{tumor} \leftarrow \text{HITL\_nnU-Net}(I_{CT})$\;
$i^* \leftarrow \operatorname{argmax}_i (|\mathcal{M}_{tumor} \cap \mathcal{M}_{vert}^{(i)}| / |\mathcal{M}_{vert}^{(i)}|)$\;

\tcp{Step 2: Topology-Guided Geometric Profiling}
$S_{post} \leftarrow \operatorname{CanalPartition}(\mathcal{M}_{vert}^{(i^*)}, \mathcal{M}_{canal}, \mathcal{M}_{tumor})$\;
$S_{collapse} \leftarrow \operatorname{ContextAwareNorm}(\mathcal{M}_{vert}^{(i^*)}, \text{Neighbors})$\;
$S_{align} \leftarrow \operatorname{KinematicProfile}(\mathcal{M}_{vert}^{(i^*)}, \text{Neighbors})$\;

\tcp{Step 3: Auxiliary Tissue \& Symptom Analysis}
$S_{bone} \leftarrow \operatorname{AnalyzeDensity}(I_{CT}, \mathcal{M}_{tumor}^{(i^*)})$\;
$S_{loc} \leftarrow \operatorname{MapLocation}(i^*)$\;
$S_{pain} \leftarrow \operatorname{LLM\_Reasoner}(T_{text})$\;

\tcp{Step 4: Aggregation \& Initial Triage}
$S_{total} \leftarrow S_{loc} + S_{pain} + S_{bone} + S_{align} + S_{collapse} + S_{post}$\;
$C_{tier} \leftarrow \operatorname{Classify}(S_{total})$\;

\tcp{Step 5: Immersive Expert Verification (Optional)}
\If{$\mathcal{V} == \mathrm{VR}$}{
    $S_{total}, C_{tier} \leftarrow \operatorname{Expert\_Correction}(\text{Profiles}, S_{total})$\;
}
\Return $S_{total}, C_{tier}$\;
\end{algorithm}

\subsection{Preprocessing and Segmentation Backbone}
Input CTs are standardized: Hounsfield Units (HU) are clipped to a bone window ($[-200, 1500]$ HU) and normalized, followed by isotropic resampling ($1.0 \times 1.0 \times 1.0$ mm$^{3}$). Vertebral body masks ($\mathcal{M}_{vert}$) are derived from TotalSegmentator v2 \cite{wasserthal2023totalsegmentator}, and the spinal canal ($\mathcal{M}_{canal}$) and tumor masks ($\mathcal{M}_{tumor}$) are generated via a 3D nnU-Net \cite{isensee2021nnu} trained on the development cohort ($N=172$). We utilized an HITL workflow, where predictions were refined by radiologists using ITK-SNAP \cite{yushkevich2006user}. This HITL validation guarantees the clinical fidelity of all anatomical targets. While the upstream segmentation employs deep learning, its outputs remain fully auditable and correctable by domain experts, ensuring that the subsequent white-box risk stratification is anchored in trustworthy anatomical ground truth.

We construct a canal-center coordinate system using principal component analysis \cite{jolliffe2016principal} on $\mathcal{M}_{canal}$. The volume is re-oriented such that the canal's principal axis aligns with the global Z-axis, eliminating positioning variability.

\subsection{Canal-Referenced Topological Partitioning}
Severe metastatic osteolysis corrupts cortical boundaries, rendering standard structural partitioning ill-posed. To resolve this, we utilize the spinal canal as a topological invariant. Let $\mathbf{v} = (x, y, z)^\top \in \mathbb{R}^3$ be voxel coordinates. We extract the canal's anterior boundary manifold, $y_{\mathrm{ant}}(z) = \min \{ y \mid \exists x \text{ s.t. } (x, y, z)^\top \in \mathcal{M}_{\mathrm{canal}} \}$, serving as a separating hyperplane. The posterior elements are rigorously defined as a half-space projection:
\begin{equation}
    \mathcal{V}_{\mathrm{post}} = \{ \mathbf{v} \in \mathcal{M}_{\mathrm{vert}} \mid \mathbf{v}_y < y_{\mathrm{ant}}(\mathbf{v}_z) \}.
\end{equation}
Consequently, tumor burden in the posterior elements is exactly $|\mathcal{V}_{\mathrm{post}} \cap \mathcal{M}_{\mathrm{tumor}}|$, evaluated against a noise threshold $\tau_V = 36 \text{ mm}^3$ derived from the voxel resolution.

\subsection{Context-Aware Morphometric Normalization}
To decouple pathological collapse from scoliotic deformity, we extract rotation-equivariant morphometry via a covariance-based Oriented Bounding Box (OBB). For vertebra $\mathcal{M}_{\mathrm{vert}}^{(i)}$ with centroid $\boldsymbol{\mu}^{(i)}$, we compute the spatial covariance matrix $\mathbf{\Sigma}^{(i)}$. Let $\mathbf{u}_1$ be its principal eigenvector, representing the true deformity-adapted craniocaudal axis. The intrinsic height is robustly derived by projecting the mask onto this axis:
\begin{equation}
    H^{(i)} = \max_{\mathbf{v}} |\mathbf{u}_1^\top (\mathbf{v} - \boldsymbol{\mu}^{(i)})|.
\end{equation}
The collapse severity $\eta = \frac{2 H^{(i)}}{H^{(i-1)} + H^{(i+1)}}$ is then calculated to establish a context-aware baseline. Discretization thresholds (e.g., $\eta < 0.50$ for severe collapse) are optimized via empirical decision boundaries on the development set.

\subsection{Kinematic Alignment Profiling}
We model the spine as a kinematic chain of OBB centroids $\mathcal{C} = \{ \boldsymbol{\mu}^{(i)} \}$. Local Kyphosis ($\theta$) is derived from the directional vectors bridging adjacent vertebrae. Let $\mathbf{d}_{\mathrm{up}} = \boldsymbol{\mu}^{(i-1)} - \boldsymbol{\mu}^{(i)}$ and $\mathbf{d}_{\mathrm{down}} = \boldsymbol{\mu}^{(i+1)} - \boldsymbol{\mu}^{(i)}$. The angular deformity is precisely calculated via the Euclidean inner product:
\begin{equation}
    \theta = \arccos \left( \frac{\mathbf{d}_{\mathrm{up}} \cdot \mathbf{d}_{\mathrm{down}}}{\|\mathbf{d}_{\mathrm{up}}\|_2 \|\mathbf{d}_{\mathrm{down}}\|_2} \right).
\end{equation}
Concurrently, translational subluxation is quantified via the 2D Intersection-over-Union (IoU) of adjacent OBBs projected onto the axial plane. Strict thresholds translate these kinematic measurements into SINS alignment scores (e.g., $S_{\mathrm{align}}=4$ for severe subluxation IoU $< 0.40$ or kyphosis $\theta > 5^{\circ}$), with threshold values determined via accuracy analysis on the development dataset.

\subsection{Integration of Standardized Clinical Modules}
While the geometric engine tackles topological ambiguities, a complete SINS profile requires integrating remaining clinical criteria. These components rely on standardized definitions rather than spatial reasoning, implemented via straightforward mapping calibrated on the development cohort:
\begin{enumerate}
    \item \textbf{Bone Lesion Quality ($S_{bone}$):} Modeled via mean HU within the tumor ROI, the lesion is classified deterministically as Lytic ($<225$ HU), Blastic ($>575$ HU), or Mixed otherwise \cite{reddington2016imaging,macedo2017bone}.
    \item \textbf{Spinal Location ($S_{loc}$):} Mapped directly from the vertebral indices to biomechanical stress zones.
    \item \textbf{Mechanical Pain ($S_{pain}$):} To maintain the system's white-box interpretability while processing unstructured clinical notes, we deploy LLMs (Qwen3, GPT-5) strictly as \textit{constrained semantic parsers} rather than black-box predictors. By enforcing Chain-of-Thought (CoT) prompting \cite{wei2022chain}, the LLM must extract explicit verbatim quotes regarding movement-exacerbated pain before outputting a binary flag. This strict evidence generation creates an instantly verifiable audit trail.
\end{enumerate}

\subsection{End-to-End Triage and Trustworthy AI} 
By aggregating the geometric innovations with auxiliary modules, TGBP functions as a comprehensive triage system. To ensure reproducibility, all pre-trained weights, thresholds, and LLM prompts will be open-sourced upon acceptance. Crucially, we reject the paradigm of uncorrectable autonomous AI. To operationalize trustworthy AI, TGBP is equipped with an immersive VR supervisory port (detailed in supplementary materials). Following our design philosophy, while TGBP operates fully automatically for high-throughput screening, this VR interface intrinsically preserves the clinician's authority to intuitively audit and intervene in the 3D geometric reasoning, maintaining ultimate clinical accountability.


\section{Experiments and Results}

\subsection{Experimental Setup and Baseline Localization}
We used multi-center, multi-cancer in-house dataset ($N=482$), and TGBP was developed and calibrated exclusively on a retrospective lung cancer cohort (GDPH1, $n=172$). Independent testing evaluated three dimensions: (i) core generalization on unseen prospective ($n=61$) and external ($n=128$) lung cohorts; (ii) cross-phenotype robustness on out-of-distribution prostate ($n=37$), breast ($n=44$), and colorectal ($n=40$) metastases; and (iii) a blinded reader study ($N=30$) comparing TGBP against five medical oncologists. Ground truth was established by two senior spine oncologists.All data acquisition and study protocols were approved by the respective institutional review boards.

\begin{table}[htbp]
\centering
\caption{Comprehensive performance of TGBP components and end-to-end system triage. Metrics show Cohen's $\kappa$ / Accuracy (\%). \textcolor{blue}{Blue}: $\kappa \geq 0.75$; \textcolor{orange}{Orange}: $0.60 \leq \kappa < 0.75$; ``-'': $\kappa$ undefined. Exact accuracy is intentionally omitted for the 18-point Total Score as ordinal agreement is optimally captured by $\kappa$.}
\label{tab:comprehensive_performance}
\resizebox{\textwidth}{!}{%
\setlength{\tabcolsep}{3pt}
\begin{tabular}{@{}l|ccc|cc|cc@{}}
\toprule
\multirow{2}{*}{\textbf{Cohort}} & \multicolumn{3}{c|}{\textbf{Geometric Components}} & \multicolumn{2}{c|}{\textbf{Auxiliary}} & \multicolumn{2}{c}{\textbf{System Triage}} \\ 
\cmidrule(lr){2-4} \cmidrule(lr){5-6} \cmidrule(lr){7-8}
 & \textbf{Collapse} & \textbf{Post. Inv.} & \textbf{Align.} & \textbf{Pain} & \textbf{Bone} & \textbf{Total} & \textbf{3-Tier} \\ 
\midrule
\rowcolor{gray!10} \multicolumn{8}{l}{\textit{Core Generalization (Lung Cancer)}} \\
Dev. Lung & \textcolor{blue}{0.785} / 76.7 & \textcolor{blue}{\textbf{0.835}} / \textbf{86.0} & 0.185 / 85.5 & \textcolor{orange}{0.653} / 81.4 & \textcolor{blue}{0.722} / 75.6 & \textcolor{orange}{0.647} & \textcolor{blue}{0.716} / 79.7 \\
Prosp. Lung & \textcolor{blue}{\textbf{0.852}} / \textbf{83.6} & \textcolor{blue}{0.796} / 88.5 & 0.228 / 91.8 & \textcolor{blue}{\textbf{0.930}} / \textbf{95.1} & \textcolor{blue}{\textbf{0.798}} / \textbf{85.2} & \textcolor{blue}{\textbf{0.788}} & \textcolor{blue}{\textbf{0.775}} / \textbf{90.2} \\
Ext. Lung & \textcolor{orange}{0.702} / 70.3 & \textcolor{blue}{0.732} / 74.2 & 0.435 / 87.5 & 0.550 / 75.8 & \textcolor{orange}{0.662} / 66.4 & 0.499 & 0.500 / 74.2 \\ 
\midrule
\rowcolor{gray!10} \multicolumn{8}{l}{\textit{Cross-Phenotype Robustness (Out-of-Distribution Metastases)}} \\
Prostate & \textcolor{blue}{\textbf{0.810}} / \textbf{91.9} & 0.494 / 73.0 & \textcolor{gray}{-} / 86.5 & 0.486 / 83.8 & \textcolor{blue}{\textbf{0.786}} / \textbf{89.2} & 0.488 & 0.560 / 64.9 \\
Breast & \textcolor{blue}{0.739} / 65.9 & \textcolor{orange}{0.640} / 75.0 & \textcolor{gray}{-} / 81.8 & \textcolor{blue}{\textbf{1.000}} / \textbf{100.0} & \textcolor{orange}{0.696} / 81.8 & 0.590 & 0.531 / 77.3 \\
Colorectal & \textcolor{blue}{\textbf{0.854}} / \textbf{82.5} & \textcolor{blue}{\textbf{0.790}} / \textbf{85.0} & 0.441 / 92.5 & \textcolor{blue}{0.812} / 87.5 & 0.598 / 80.0 & \textcolor{blue}{\textbf{0.771}} & \textcolor{blue}{\textbf{0.700}} / \textbf{87.5} \\ 
\bottomrule
\end{tabular}%
}
\end{table}

Crucially, downstream TGBP profiling operated on fully automated anatomical parsing from frozen segmentation models (HITL was restricted to the development set). Omitting manual voxel refinement intentionally subjects our geometric engine to real-world segmentation variance, verifying true end-to-end viability.

\subsection{System-Level Triage and Component Evaluation}
As detailed in Table \ref{tab:comprehensive_performance}, the topology-guided geometric engine reliably extracted structural profiles across varied osteolytic conditions. Canal-referenced partitioning handled severe cortical destruction robustly (e.g., maintaining substantial agreement, $\kappa=0.790$, in aggressive colorectal metastases). Covariance-based morphometry validated its scoliotic invariance by achieving near-perfect agreement in prospective lung ($\kappa=0.852$) and osteoblastic prostate ($\kappa=0.810$) cohorts. While scanner heterogeneity in the multi-center external cohort slightly degraded HU-dependent bone lesion classification, TGBP successfully integrated all components to achieve highly reliable end-to-end 3-tier triage (up to 90.2\% accuracy).

To justify our decoupled white-box approach, we evaluated end-to-end classification baselines (ResNet\cite{targ2016resnet}/DenseNet\cite{huang2018condensenet}) for direct 3-tier classification (Table \ref{tab:dl_baseline}). Because standard SINS incorporates mechanical pain (up to 3 points) from unstructured clinical texts, strict clinical thresholds inherently penalize these single modality models. To guarantee a fair, imaging-only evaluation, we shifted the baselines' ground-truth classification boundaries downward by exactly 3 points (the maximum pain score). Even with this favorable calibration isolating their structural assessment capability, the best baseline (DenseNet-121, $\kappa=0.582$) underperformed TGBP, highlighting the fundamental inadequacy of black-box paradigms in resolving complex topological ambiguity.

\subsection{Bridging the Diagnostic Gap: The Compounding Error}
We quantified the vulnerability of manual opportunistic screening via a reader study ($N=30$, Table \ref{tab:reader_study}). A catastrophic diagnostic gap emerged: while medical oncologists achieved moderate agreement on basic symptoms, they struggled severely with spatial geometry (mean $\kappa=0.570$ for posterolateral involvement). Consequently, individual variance cascaded into near-random Total SINS agreement (mean $\kappa=0.207$). TGBP's deterministic topology rules prevented this compounding failure, elevating total score agreement to $\kappa=0.625$ and 3-tier triage to $\kappa=0.803$, democratizing expert-level structural assessment.

\begin{table}[htbp] 
\centering
\caption{Comparison with end-to-end deep learning baselines on out-of-distribution cohorts. $^\dagger$To ensure a fair imaging-only evaluation, the ground-truth 3-tier classification thresholds for these pure vision models were shifted downward by exactly 3 points (the maximum score of the omitted mechanical pain component).}
\label{tab:dl_baseline}
\scriptsize
\setlength{\tabcolsep}{12pt}
\begin{tabular}{@{}llc@{}}
\toprule
\textbf{Methodology} & \textbf{Architecture} & \textbf{3-Tier Class ($\kappa$)} \\
\midrule
\multirow{4}{*}{Black-Box Baseline$^\dagger$} 
 & ResNet-50 & 0.377 \\
 & ResNet-152 & 0.413 \\
 & DenseNet-169 & 0.525 \\
 & DenseNet-121 & 0.582 \\
\midrule
\textbf{Proposed (White-Box)} & \textbf{TGBP (Aggregated)} & \textbf{0.597} \\ 
\bottomrule
\end{tabular}
\end{table}

\begin{table}[htbp] 
\centering
\caption{Quantification of the Diagnostic Gap ($N=30$). TGBP robustly prevents the compounding error in manual Total Score estimation. \textcolor{blue}{Blue}: $\kappa \geq 0.75$.}
\label{tab:reader_study}
\scriptsize
\setlength{\tabcolsep}{5pt}
\begin{tabular}{@{}lcccc@{}}
\toprule
\textbf{SINS Component / Output} & \textbf{TGBP ($\kappa$)} & \textbf{Oncologists (Mean $\kappa$)} & \textbf{Range} & \textbf{$\Delta\kappa$} \\ 
\midrule
Spinal Location & \textcolor{blue}{\textbf{1.000}} & 0.485 & 0.35--0.60 & \textcolor{teal}{+0.515} \\
Bone Lesion Quality & \textcolor{blue}{\textbf{0.806}} & 0.421 & 0.35--0.49 & \textcolor{teal}{+0.385} \\
Mechanical Pain & \textcolor{blue}{\textbf{0.784}} & 0.680 & 0.52--0.77 & \textcolor{teal}{+0.104} \\
Vertebral Collapse & \textcolor{blue}{\textbf{0.858}} & 0.762 & 0.67--0.84 & \textcolor{teal}{+0.096} \\
Posterolateral Inv. & \textcolor{blue}{\textbf{0.857}} & 0.570 & 0.47--0.66 & \textcolor{teal}{+0.287} \\
Spinal Alignment & 0.277 & -0.006 & -0.13--0.06 & +0.283 \\ 
\midrule
\textbf{Total SINS Score} & 0.625 & 0.207 & 0.04--0.42 & \textcolor{red}{\textbf{+0.418}} \\
\textbf{3-Tier Stability Class} & \textcolor{blue}{\textbf{0.803}} & 0.565 & 0.45--0.70 & \textcolor{red}{\textbf{+0.238}} \\
\bottomrule
\end{tabular}
\end{table}

\section{Discussion and Conclusion}
We proposed TGBP, a white-box framework that fundamentally bridges the diagnostic gap in opportunistic screening for spinal instability. By decoupling anatomical perception from explicit structural reasoning, TGBP resolves the topological ambiguity of metastatic osteolysis via canal-referenced partitioning and extracts rotation-equivariant morphometry. Our extensive validation ($N=482$) demonstrates that anchoring SINS assessment on transparent geometry achieves expert-level 3-tier triage (90.2\% accuracy) and prevents the catastrophic compounding errors inherent in manual evaluation ($\kappa=0.625$ vs.\ $0.207$). 

Despite its robustness, TGBP has intrinsic limitations. First, our topological partitioning fundamentally relies on the spinal canal as a geometric anchor—a prior that may fail in cases of massive epidural tumor extension causing total canal obliteration. Second, while OBB-based morphometry provides a validated structural proxy, it derives from static, non-weight-bearing supine CTs, inherently oversimplifying the anisotropic biomechanical failure of trabecular micro-architecture. 

TGBP operationalizes trustworthy AI principles by integrating an immersive VR supervisory port for essential expert verification. Future work will incorporate dynamic finite element analysis (FEA) and multimodal foundation models to overcome these static limitations, ultimately democratizing proactive, expert-level surgical triage in routine oncology and ensuring critical prophylactic windows are no longer missed.

\bibliographystyle{splncs04}
\bibliography{reference}

@article{pickhardt2020automated,
  title={Automated {CT} biomarkers for opportunistic prediction of future cardiovascular events and mortality in an unselected screening population: a retrospective cohort study},
  author={Pickhardt, Perry J and Graffy, Peter M and others},
  journal={The Lancet Digital Health},
  volume={2},
  number={4},
  pages={e192--e200},
  year={2020},
  publisher={Elsevier}
}

@article{cole2008metastatic,
  title={Metastatic epidural spinal cord compression},
  author={Cole, John S and Patchell, Roy A},
  journal={The Lancet Neurology},
  volume={7},
  number={5},
  pages={459--466},
  year={2008},
  publisher={Elsevier}
}

@article{fisher2010novel,
  title={A novel classification system for spinal instability in neoplastic disease: an evidence-based approach and expert consensus from the {Spine Oncology Study Group}},
  author={Fisher, Charles G and DiPaola, Christian P and Ryken, Timothy C and others},
  journal={Spine},
  volume={35},
  number={22},
  pages={E1221--E1229},
  year={2010},
  publisher={LWW}
}

@inproceedings{hu2019topology,
  title={Topology-preserving deep image segmentation},
  author={Hu, Xiaoling and Li, Fuxin and Samaras, Dimitris and Chen, Chao},
  booktitle={Advances in Neural Information Processing Systems},
  volume={32},
  pages={5658--5669},
  year={2019}
}

@article{wasserthal2023totalsegmentator,
  title={{TotalSegmentator}: Robust bounding box detection and 104 anatomic structures segmentation in {CT}},
  author={Wasserthal, Jakob and Breit, Hanns-Christian and Meyer, Manfred T and others},
  journal={Radiology: Artificial Intelligence},
  volume={5},
  number={5},
  pages={e230024},
  year={2023},
  publisher={Radiological Society of North America}
}

@article{peng2024study,
  title={A study on whether deep learning models based on {CT} images for bone density classification and prediction can be used for opportunistic osteoporosis screening},
  author={Peng, Tao and Zeng, Xiaohui and Li, Yang and Li, Man and Pu, Bingjie and Zhi, Biao and Wang, Yongqin and Qu, Haibo},
  journal={Osteoporosis International},
  volume={35},
  number={1},
  pages={117--128},
  year={2024},
  publisher={Springer}
}

@article{isensee2021nnu,
  title={{nnU-Net}: a self-configuring method for deep learning-based biomedical image segmentation},
  author={Isensee, Fabian and Jaeger, Paul F and Kohl, Simon AA and Petersen, Jens and Maier-Hein, Klaus H},
  journal={Nature Methods},
  volume={18},
  number={2},
  pages={203--211},
  year={2021},
  publisher={Nature Publishing Group}
}

@article{yushkevich2006user,
  title={User-guided {3D} active contour segmentation of anatomical structures: significantly improved efficiency and reliability},
  author={Yushkevich, Paul A and Piven, Joseph and Hazlett, Heather Cody and others},
  journal={Neuroimage},
  volume={31},
  number={3},
  pages={1116--1128},
  year={2006},
  publisher={Elsevier}
}

@article{jolliffe2016principal,
  title={Principal component analysis: a review and recent developments},
  author={Jolliffe, Ian T and Cadima, Jorge},
  journal={Philosophical Transactions of the Royal Society A: Mathematical, Physical and Engineering Sciences},
  volume={374},
  number={2065},
  pages={20150202},
  year={2016},
  publisher={The Royal Society Publishing}
}

@article{chan2025llm,
  title={Large Language Model ({LLM})-Predicted and {LLM}-Assisted Calculation of the {Spinal Instability Neoplastic Score} ({SINS}) Improves Clinician Accuracy and Efficiency},
  author={Chan, Matthew Ding Zhou and Tjio, Calvin Kai En and Chan, Tammy Li Yi and Tan, Yi Liang and Chua, Alynna Xu Ying and Loh, Sammy Khin Yee and Leow, Gabriel Zi Hui and Gan, Ming Ying and Lim, Xinyi and Choo, Amanda Kexin and others},
  journal={Cancers},
  volume={17},
  number={19},
  pages={3198},
  year={2025},
  publisher={MDPI}
}

@article{ma2023medsam,
  title={Segment anything in medical images},
  author={Ma, Jun and He, Yuting and Li, Feifei and Han, Lin and You, Chenyu and Wang, Bo},
  journal={Nature communications},
  volume={15},
  number={1},
  pages={654},
  year={2024},
  publisher={Nature Publishing Group UK London}
}

@article{liu2024learning,
  title={Learning-based bone quality classification method for spinal metastasis},
  author={Peng, Shiqi and Lai, Bolin and Yao, Guangyu and Zhang, Xiaoyun and Zhang, Ya and Wang, Yan-Feng and Zhao, Hui},
  booktitle={International Workshop on Machine Learning in Medical Imaging},
  pages={426--434},
  year={2019},
  organization={Springer}
}

@article{chow2007palliative,
  title={Palliative radiotherapy trials for bone metastases: a systematic review},
  author={Chow, Edward and Harris, Kristin and Fan, Grace and Tsao, May and Sze, Wai M},
  journal={Journal of Clinical Oncology},
  volume={25},
  number={11},
  pages={1423--1436},
  year={2007},
  publisher={American Society of Clinical Oncology}
}

@article{bartels2008spinal,
  title={Spinal extradural metastasis: review of current treatment options},
  author={Bartels, Ronald HMA and van der Linden, Yvette M and van der Graaf, Winette TA},
  journal={CA: a cancer journal for clinicians},
  volume={58},
  number={4},
  pages={245--259},
  year={2008},
  publisher={Wiley Online Library}
}

@article{costelloe2009imaging,
  title={Imaging bone metastases in breast cancer: techniques and recommendations for diagnosis},
  author={Costelloe, Colleen M and Rohren, Eric M and Madewell, John E and Hamaoka, Tsuyoshi and Theriault, Richard L and Yu, Tse-Kuan and Lewis, Valerae O and Ma, Jingfei and Stafford, R Jason and Tari, Ana M and others},
  journal={The lancet oncology},
  volume={10},
  number={6},
  pages={606--614},
  year={2009},
  publisher={Elsevier}
}

@inproceedings{wen2025topology,
  title={Topology-preserving image segmentation with spatial-aware persistent feature matching},
  author={Wen, Bo and Zhang, Haochen and Bartsch, Dirk-Uwe G and Freeman, William and Nguyen, Truong and An, Cheolhong},
  booktitle={Proceedings of the IEEE/CVF International Conference on Computer Vision},
  pages={5762--5771},
  year={2025}
}

@article{ye2026unveiling,
  title={Unveiling and Bridging the Functional Perception Gap in {MLLMs}: Atomic Visual Alignment and Hierarchical Evaluation via {PET-Bench}},
  author={Ye, Zanting and Niu, Xiaolong and Wu, Xuanbin and Han, Xu and Liu, Shengyuan and Hao, Jing and Peng, Zhihao and Sun, Hao and Lv, Jieqin and Wang, Fanghu and others},
  journal={arXiv preprint arXiv:2601.02737},
  year={2026}
}

@article{jiang2025hulu,
  title={{Hulu-med}: A transparent generalist model towards holistic medical vision-language understanding},
  author={Jiang, Songtao and Wang, Yuan and Song, Sibo and Hu, Tianxiang and Zhou, Chenyi and Pu, Bin and Zhang, Yan and Yang, Zhibo and Feng, Yang and Zhou, Joey Tianyi and others},
  journal={arXiv preprint arXiv:2510.08668},
  year={2025}
}

@article{guo2024radiographic,
  title={Radiographic imaging and diagnosis of spinal bone tumors: {AlexNet} and {ResNet} for the classification of tumor malignancy},
  author={Guo, Chengquan and Chen, Yan and Li, Jianjun},
  journal={Journal of bone oncology},
  volume={48},
  pages={100629},
  year={2024},
  publisher={Elsevier}
}

@article{reddington2016imaging,
  title={Imaging characteristic analysis of metastatic spine lesions from breast, prostate, lung, and renal cell carcinomas for surgical planning: Osteolytic versus osteoblastic},
  author={Reddington, Justin A and Mendez, Gustavo A and Ching, Alex and Dai Kubicky, Charlotte and Klimo Jr, Paul and Ragel, Brian T},
  journal={Surgical neurology international},
  volume={7},
  number={Suppl 13},
  pages={S361},
  year={2016}
}

@article{macedo2017bone,
  title={Bone metastases: an overview},
  author={Macedo, Filipa and Ladeira, Katia and Pinho, Filipa and Saraiva, Nadine and Bonito, Nuno and Pinto, Lu{\'\i}sa and Gon{\c{c}}alves, Francisco},
  journal={Oncology reviews},
  volume={11},
  number={1},
  pages={321},
  year={2017}
}

@article{wei2022chain,
  title={Chain-of-thought prompting elicits reasoning in large language models},
  author={Wei, Jason and Wang, Xuezhi and Schuurmans, Dale and Bosma, Maarten and Xia, Fei and Chi, Ed and Le, Quoc V and Zhou, Denny and others},
  journal={Advances in neural information processing systems},
  volume={35},
  pages={24824--24837},
  year={2022}
}

@article{targ2016resnet,
  title={{ResNet} in {ResNet}: Generalizing residual architectures},
  author={Targ, Sasha and Almeida, Diogo and Lyman, Kevin},
  journal={arXiv preprint arXiv:1603.08029},
  year={2016}
}

@inproceedings{huang2018condensenet,
  title={{CondenseNet}: An efficient {DenseNet} using learned group convolutions},
  author={Huang, Gao and Liu, Shichen and Van der Maaten, Laurens and Weinberger, Kilian Q},
  booktitle={Proceedings of the IEEE conference on computer vision and pattern recognition},
  pages={2752--2761},
  year={2018}
}

@article{fourney2011spinal,
  title={{Spinal Instability Neoplastic Score}: an analysis of reliability and validity from the spine oncology study group},
  author={Fourney, Daryl R and Frangou, Evan M and Ryken, Timothy C and DiPaola, Christian P and Shaffrey, Christopher I and Berven, Sigurd H and Bilsky, Mark H and Harrop, James S and Fehlings, Michael G and Boriani, Stefano and others},
  journal={Journal of clinical oncology},
  volume={29},
  number={22},
  pages={3072--3077},
  year={2011},
  publisher={American Society of Clinical Oncology}
}

@article{xu2025lingshu,
  title={{Lingshu}: A generalist foundation model for unified multimodal medical understanding and reasoning},
  author={Xu, Weiwen and Chan, Hou Pong and Li, Long and Aljunied, Mahani and Yuan, Ruifeng and Wang, Jianyu and Xiao, Chenghao and Chen, Guizhen and Liu, Chaoqun and Li, Zhaodonghui and others},
  journal={arXiv preprint arXiv:2506.07044},
  year={2025}
}

@article{sellergren2025medgemma,
  title={{MedGemma} technical report},
  author={Sellergren, Andrew and Kazemzadeh, Sahar and Jaroensri, Tiam and Kiraly, Atilla and Traverse, Madeleine and Kohlberger, Timo and Xu, Shawn and Jamil, Fayaz and Hughes, C{\'\i}an and Lau, Charles and others},
  journal={arXiv preprint arXiv:2507.05201},
  year={2025}
}

\end{document}